\documentclass[12pt]{aastex}

\usepackage{natbib}
\usepackage{graphicx}

\usepackage{times}

\title{A Giant Planet Around a Metal-poor Star of Extragalactic Origin}

\author
{Johny Setiawan\altaffilmark{1}, Rainer J.~Klement\altaffilmark{1}, Thomas Henning\altaffilmark{1}, Hans-Walter Rix\altaffilmark{1},\\
Boyke Rochau\altaffilmark{1}, Jens Rodmann\altaffilmark{2}, Tim Schulze-Hartung\altaffilmark{1}}

\altaffiltext{1}{Max-Planck-Institut f\"ur Astronomie, K\"onigstuhl 17, D-69117 Heidelberg; setiawan@mpia.de}
\altaffiltext{2}{European Space Agency, Space Environment and Effects Section, ESTEC, Keplerlaan 1, 2201 AZ Noordwijk, The Netherlands}

\date{}

\begin{document} 


\baselineskip24pt


\maketitle


\begin{abstract}
Stars in their late stage of evolution, such as Horizontal Branch stars, are still largely unexplored for planets. 
We report the detection of a planetary companion around \mbox{HIP 13044}, 
a very metal-poor star on the red Horizontal Branch, 
based on radial velocity observations with a high-resolution spectrograph 
at the 2.2-m MPG/ESO telescope. The star's periodic radial velocity 
variation of $P=16.2$ days caused by the planet can be distinguished
from the periods of the stellar activity indicators.
The minimum mass of the planet is 1.25 M$_\mathrm{jup}$ and its orbital semi-major axis 0.116 AU.
Because \mbox{HIP 13044} belongs to a group of stars that have been accreted from a disrupted satellite galaxy of
the Milky Way, the planet most likely has an extragalactic origin.
\end{abstract}

\section{Main text}
In the last two decades, several hundred planets have been detected beyond our Solar-system. 
Most of these extra-solar planets orbit sun-like stars. 
A small number have been detected around stars that are in their late evolutionary state, such as Red 
Giant Branch (RGB) stars and pulsars. The phase directly after the 
RGB stage, the Horizontal Branch (HB), however, is still unexplored; 
therefore, there is no empirical evidence for whether close-in planets, 
i.e., those with semi-major axes less than 0.1 AU, survive 
the giant phase of their host stars.

Besides its evolutionary stage, a star's chemical composition appears to 
be a major indicator of its probability for hosting a planet. 
Previous studies, e.g., \citet{gon97}, showed that main-sequence (MS) stars 
that host giant planets are metal-rich. 
This finding is supported by the large exoplanet search surveys around MS stars 
reporting a connection
between planet frequency and metallicity \citep{san04,val05}, 
and a survey of 160 metal-poor main-sequence stars
finding no evidence for Jovian planets \citep{soz09}. 

Until now, only very few planets have been detected around stars 
with metallicities as low as [Fe/H]=\, $-1$, i.e. 10\% of the sun's metallicity. 
The detection of PSR B1620 b, a Jovian planet orbiting a pulsar 
in the core of the metal-poor globular cluster M4 ([Fe/H]=$-1.2$), 
suggests, however, that planets may form around 
metal-poor stars \citep{ford00,sig03}, although the formation mechanism of this particular planet might be linked to the dense cluster environment
\citep{bee04}.

We used the Fibre-fed Extended Range Optical Spectrograph (FEROS), 
a high-resolution spectrograph ($R=48,000$) 
attached to the 2.2 meter Max-Planck Gesellschaft/European Southern 
Observatory (MPG/ESO) telescope\footnote{The observations of \mbox{HIP 13044} were carried out from September 2009 
until July 2010. The spectrograph covers a wavelength range from 350 to 920 nm \citep{kau00}.
To measure the RV values of \mbox{HIP 13044} we used a cross-correlation technique, where the
stellar spectrum is cross-correlated with a numerical template (mask) designed for stars of the spectral type F0 
(Section~\ref{text2}).}, to observe the star \mbox{HIP 13044}. 
This star  is classified as a red HB (RHB) star (Fig.~1) 
and its metal content is [Fe/H]$_\mathrm{mean}=-2.09$ \citep{bee90,chi00,car08b,roe10}, 
i.e. about 1\% that of the Sun.
So far, \mbox{HIP 13044} is not known as a binary system. 
Detailed stellar parameters can be found in Supporting Online Material, Section~\ref{text1}.
 
Previous radial velocity (RV) measurements of \mbox{HIP 13044} showed 
a systematic velocity of about 300 $\mathrm{km\,s}^{-1}$ with respect to the Sun, 
indicating that the star belongs to the stellar halo \citep{car86}. 
Indeed, the star has been connected 
to the Helmi stream \citep{hel99}, a group of stars that share similar orbital parameters that
stand apart from those of the bulk of other stars in the solar neighborhood. The Helmi stream stars move on prograde eccentric 
orbits ($R_\mathrm{peri}\sim7$ kpc, $R_\mathrm{apo}\sim16$ kpc) that
reach distances up to $\vert z\vert_\mathrm{max}\sim13$ kpc above and below
the galactic plane. 
From that, it has been concluded that these stars were once bound to a satellite galaxy of the Milky Way \citep{hel99,chi00}
that was tidally disrupted 6--9 Ga ago \citep{kep07}. 
 
The variation of the RV between our observations at different epochs has a 
semi-amplitude of 120 $\mathrm{m\,s}^{-1}$\footnote{In order to search for periodic variations, we used periodogram analysis techniques, 
which are capable of treating missing values and unevenly spaced time points.} (Fig. 2).
The Generalized Lomb Scargle (GLS) periodogram \citep{zec09} 
reveals a significant RV periodicity at $P=16.2$ days with 
a False Alarm Probability of $5.5\times10^{-6}$. 
Additional analysis, using a Bayesian algorithm \citep{gre05}, yields
a similar period around 16 days.
Such RV variation can be induced by an unseen orbiting companion, 
by moving/rotating surface inhomogeneities or by non-radial 
stellar pulsations. Exploring both stellar rotational modulation 
and pulsations is critical when probing the
presence of a planetary companion, because they can produce a similar or even
the same RV variation, mimicking a Keplerian motion.

A well-established technique to detect stellar rotational modulation 
is to investigate the line profile asymmetry or bisector \citep{gray08} and Ca II lines 
(Section~\ref{text3}). 
Surface inhomogeneities, such as starspots 
and large granulation cells, produce asymmetry in the spectral line
profiles. When a spotted star rotates, the barycenter of the line profiles moves
periodically and the variation can mimic a RV variation caused by an orbiting
companion. Instead of measuring the bisectors, one can equivalently use 
the bisector velocity spans (BVS) to search for rotational 
modulation \citep{hat96}.
Adopting this technique, we have measured BVS from the stellar spectra. 
There is only a weak correlation between BVS and RV (correlation
coefficient =$-0.13$), but the BVS variation shows a clear periodicity 
with $P=$5.02 days (Section~\ref{sec:bisector}). 
No period around 16 days is found in the BVS variation.  

In addition to the BVS analysis, we investigated the variation of the 
Ca II $\lambda$849.8 line, which is one of the Ca II infrared triplet lines. 
From the observed Ca II 849.8 equivalent-width variations we computed 
a mean period of 6.05 days (Section~\ref{ew}), 
which is in the same order of the period of the BVS variation. 
We adopted the mean period of both stellar activity indicators, 
$P_{\mathrm{rot}}=5.53\pm0.73$ days, 
as the stellar rotation period of \mbox{HIP 13044} and then calculated the 
inclination angle of the stellar rotation axis, 
which follows from $ P_{\mathrm{rot}}/\sin i = 2\pi R_* / v \sin i$. 
With a stellar radius 
$R_*=6.7\,\mathrm{R}_\odot$ \cite{car08a} and our adopted 
value for the projected rotational velocity, 
$v \sin i$=10.25 $\mathrm{km\,s}^{-1}$, which was derived from the observed line broadening 
(Section~\ref{text1}), we obtained an inclination angle $i= 9.7\pm1.3$ deg.
Thus, the real stellar rotation velocity is 
$\sim$62 $\mathrm{km\,s}^{-1}$, which is typical for an early F type MS-star 
but relatively high for HB stars. 

An explanation for this high rotation velocity is the assumption that \mbox{HIP 13044} 
has engulfed its close-in planets during the red giant phase. 
Infalling planets are able to spin-up their host star \citep{sok98,lev09,car09}, and this mechanism has been suggested to explain the
high $v \sin i$ values observed for many RGB and HB stars \citep{car03}.

We observed variations of \mbox{HIP 13044} 
in the photometric data from the Hipparcos satellite \citep{per97} 
and SuperWASP \citep{pol06} (Section~\ref{phot}). 
While the Hipparcos data shows only a marginal significant 
periodicity of 7.1 hours (FAP=1.8\%), the SuperWASP data 
shows few intra-day periodicities with FAP$\sim$1\% and 
two significant periodicities at 1.39 (FAP=$5\times10^{-4}$) 
and 3.53 days (FAP=$2\times10^{-4}$). These two periods, 
however, are most likely harmonic to each other ($1.4^{-1}+3.5^{-1}=1$).
It is expected that \mbox{HIP 13044} oscillates only 
at pulsationally unstable overtones of high order \citep{xio98}. 
Observations of one RHB star in the metal-poor 
globular cluster NGC 6397 \citep{ste09} as well as 
theoretical predictions \citep{xio98} set these 
periods in the range of a few hours to a day or so. 
No clear theoretical predictions for a star with parameters 
similar to \mbox{HIP 13044} exist, hence it is possible 
that some high-order oscillations are able to explain the 1.4 or 3.5 
day signal. What is important, however, is that there is 
no signal of a period around 16.2 d in the photometric data.

The arguments above show that neither stellar rotational modulation 
nor pulsations are plausible sources of the observed periodic 
RV variation. Therefore, the best explanation for the 
$\sim$16 days period is the presence of an unseen companion. 
We computed its orbital solution (Table~1).
Its minimum mass lies securely in the planetary mass domain, 
even with a plausible $\sin{i}$ uncertainty. 
With an eccentricity of 0.25 and a semi-major axis of 0.116 AU, the planet
has a periastron distance of about 0.087 AU which is $\approx$2.8 times 
the present stellar radius. 
The periastron is $\sim$0.06 AU away from the stellar surface.

\begin{table}[ht]
 \begin{center}
 \caption{Orbital parameters of \mbox{HIP 13044} b}
   \begin{tabular}{lll}
 \hline
 \hline 
 $P$      	    &  16.2    $\pm$ 0.3 	    & days  \\
 $K_{1}$            &  119.9   $\pm$ 9.8	    & $\mathrm{m\,s}^{-1}$\\
 $e$                &  0.25    $\pm$ 0.05	    &	    \\
 $\omega$           &  219.8   $\pm$ 1.8	    & deg   \\
 $JD_{0}-2450000$   &  5109.78 $\pm$ 0.02	    & days  \\
 $\chi^{2}$ 	    &  32.35 		    	    & $\mathrm{m\,s}^{-1}$\\
 $rms$         	    &  50.86     		    & $\mathrm{m\,s}^{-1}$\\
 $m_{2} sin{i}$     &  1.25    $\pm$ 0.05	   & M$_{\mathrm{Jup}}$ \\
 $a$	            &  0.116   $\pm$ 0.01	   & AU    \\
 \hline 
 \hline
 
 \end{tabular}
 \end{center}
\end{table}

Because a large number of known exoplanets have orbital semi-major 
axes between 0.01 and 0.06 AU, the distance between the periastron and
the star \mbox{HIP 13044} itself is not unusual.
The non-circular orbit ($e=$0.25), however, is not expected for 
a close-in giant planet around a post RGB star. 

In the case of \mbox{HIP 13044}, the original orbit could have been disturbed or 
changed during the evolution of 
the star-planet-system, in particular during the RGB phase \citep{sok98}. 
Interestingly, the orbital period of \mbox{HIP 13044} b is close to three times
the stellar rotation period. There are a number of known planetary systems 
which also have such a ``coupling'' between the stellar rotation and orbital periods,
e.g. Tau Boo (1:1), HD 168433 (1:2), HD 90156 (1:2) and HD 93083 (1:3).
Such planetary systems are particularly interesting to study 
star-planet interactions \citep{shk08}.

So far, there are only very few planet or brown dwarf 
detections around post RGB stars besides the pulsar planets, 
namely \mbox{V391 Peg} \citep{sil07}, \mbox{HW Vir} \citep{lee09} 
and \mbox{HD 149382} \citep{gei09} (Fig.~1). These are, however, substellar companions around 
subdwarf-B or Extreme Horizontal Branch (EHB) stars, i.e., 
the nature of their host stars differs from that of \mbox{HIP 13044}, an RHB star. 
Contrary to RGB stars, such as G and K giants \citep{wal92,hat93,set04,doe07} and 
subgiants \citep[e.g.][]{joh10}, HB stars have not been yet extensively surveyed for planets.

While at least 150 main-sequence stars are known to bear close-in ($a=0.1$
AU) giant planets, so far no such planets have been reported
around RGB stars. A possible explanation is that the inner planets have
been engulfed by the star when the stellar atmosphere expanded during the
giant phase. The survival of \mbox{HIP 13044 b} during that phase is theoretically possible 
under certain circumstances \citep{liv83,sok98,bea10}. 
It is also possible that the planet's orbit decayed through tidal interaction 
with the stellar envelope.
However, a prerequisite to survival is then that the mass 
loss of the star stops before the planet would have been evaporated or accreted. 
Assuming asymmetric mass loss, velocity kicks could have increased 
the eccentricity of \mbox{HIP 13044 b} to its current, somewhat high value \citep{hey07}. 
The same could be achieved by interaction with a third body in the system.

Interestingly, a survey to characterize the multiplicity of EHB stars showed that more than
60\% of the sample are close binaries \citep{max01}. Their orbital radii are much smaller
than the stellar radius in the RGB phase. This could be explained by the high friction 
in the interstellar medium, which would move a distant companion towards the primary. 
Such spiral-in mechanism could also take place in the RGB-to-RHB transition phase.
Similar to the binary case, a distant giant planet in the
RGB phase can move towards the primary into a smaller orbit. 
Consequently, the resulting close orbiting planets will be engulfed 
when the stellar envelope expands again in the next giant phase. 
\mbox{HIP 13044} b could be a planet that is just about to be engulfed by its star. 

\mbox{HIP 13044}, with a mean metallicity estimate of [Fe/H]=-2.1, is far more 
metal-poor than any previously known exoplanet hosting star (Fig.~3). 
For the existing theories of hot giant planet formation, 
metallicity is an important parameter: in particular, it is fundamental for the
core-accretion planet formation model \citep{ida05}.  
It might be that initially, in the planet formation phase, \mbox{HIP 13044} 
had a higher metallicity, and that during its subsequent evolution, it lost its 
heavier elements.
For example, during the giant phase, heavy elements could have had been incorporated 
into dust grains and then separated from the star's atmosphere \citep{mat92}.
However, given the star's membership to the Helmi stream, in which
the most metal-rich sub-dwarfs known so far have [Fe/H]$\sim-1.5$ 
\citep{kle09}, we do not expect its initial Fe abundance to exceed this value. 

Finally, as a member of the Helmi stream, \mbox{HIP 13044} most probably 
has an extragalactic origin. This implies that its history is likely different from 
those of the majority of known planet-hosting stars. 
\mbox{HIP 13044} was probably
attracted to the Milky Way several Ga ago. Before that, it
could have had belonged to a satellite galaxy of the Milky Way similar to
Fornax or the Sagittarius dwarf spheroidal galaxy \citep{hel99}. 
Because of the long galactic relaxation timescale, it is extremely unlikely that \mbox{HIP 13044} b
joined its host star through exchange with some Milky Way star, after the
former had been tidally stripped. The planet \mbox{HIP 13044} b could thus have a non-Galactic origin.

 \begin{figure}
 \includegraphics[width=16cm]{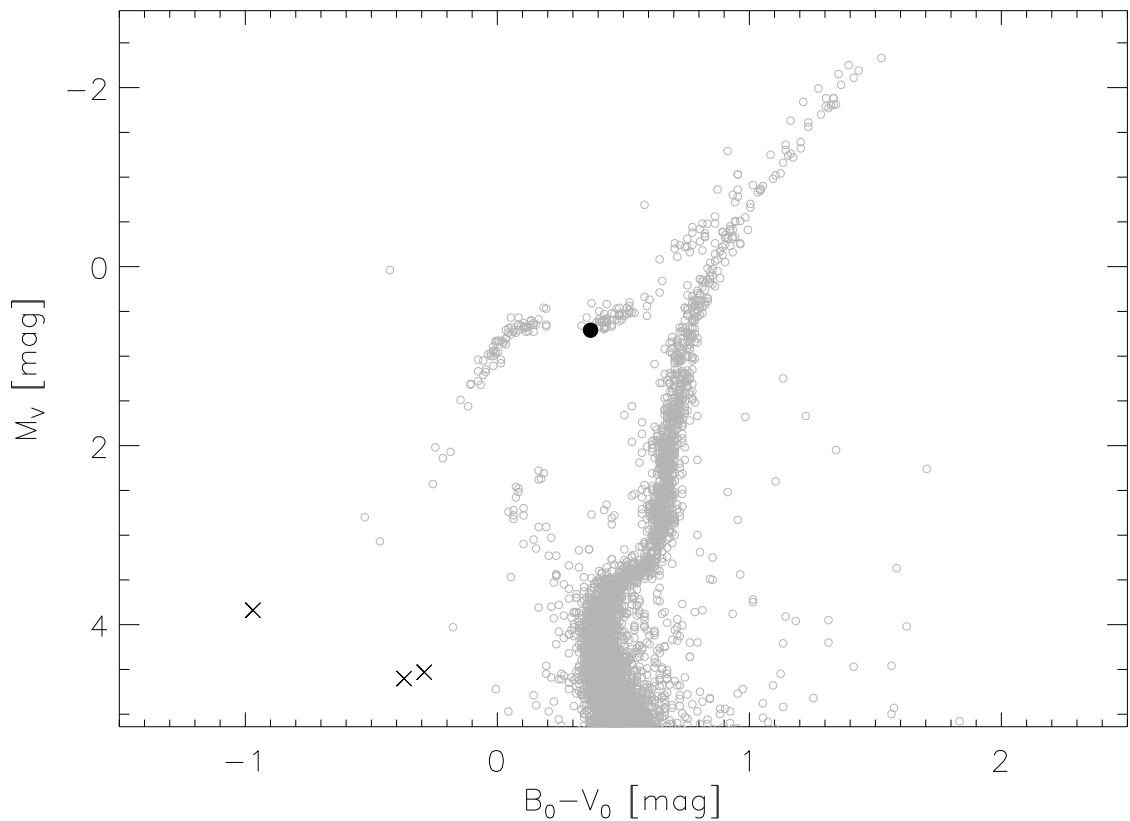}
 \caption{Location of \mbox{HIP 13044} in a M$_V$ vs. $B-V$ color-magnitude 
 diagram (CMD) shown as a black dot superimposed to the CMD of Messier 3 (grey open circles) 
 based on the photometry by \citet{buo94}. Apparent magnitudes have been converted to 
 absolute magnitudes by considering the distance modulus and extinction given by 
 \citet{har96}. The gap separating the blue and red Horizontal Branch (HB) is due to 
 RR Lyrae instability strip. The CMD location of \mbox{HIP 13044} implies that it is a core He-burning 
 star, located a the blue edge of the RHB. Further candidates for post RGB stars hosting 
 planets/brown dwarf, V391 Peg, HW Vir and HD 149382 \citep{sil07,lee09,gei09} are 
 displayed as crosses.}
 \end{figure}

 \begin{figure*}
 \includegraphics[width=16cm]{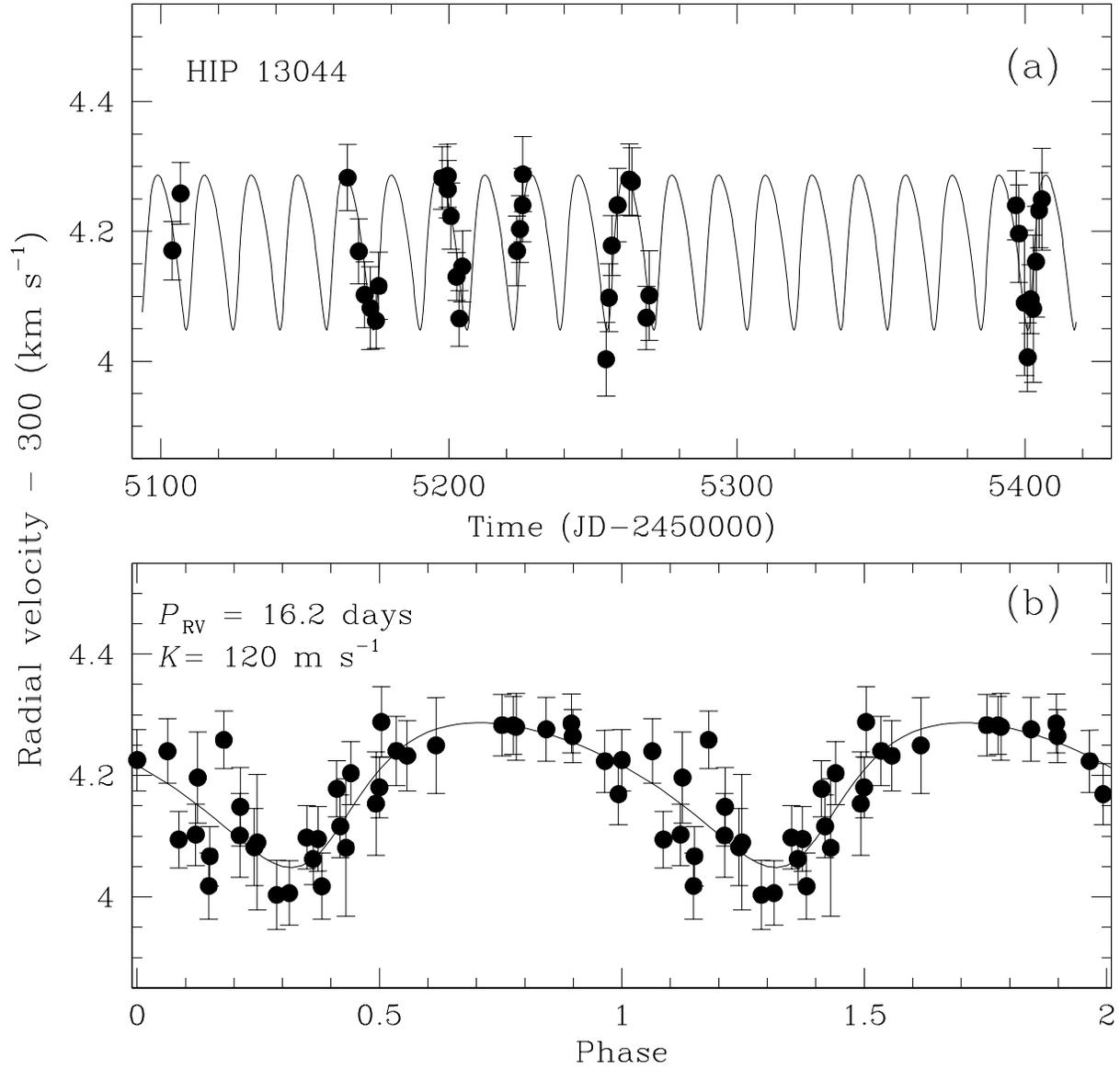}
 \caption{(Upper panel) RV variation of \mbox{HIP 13044}. 
 The RV values have been computed from the mean RVs of 20 usable echelle 
 orders of the individual spectrum. The error bars have been calculated 
 from the standard error of the mean RV of each order.
 (Lower panel) RV variation phase-folded with $P=$16.2 days.}
 \end{figure*}

 \begin{figure*}
 \includegraphics[width=16cm]{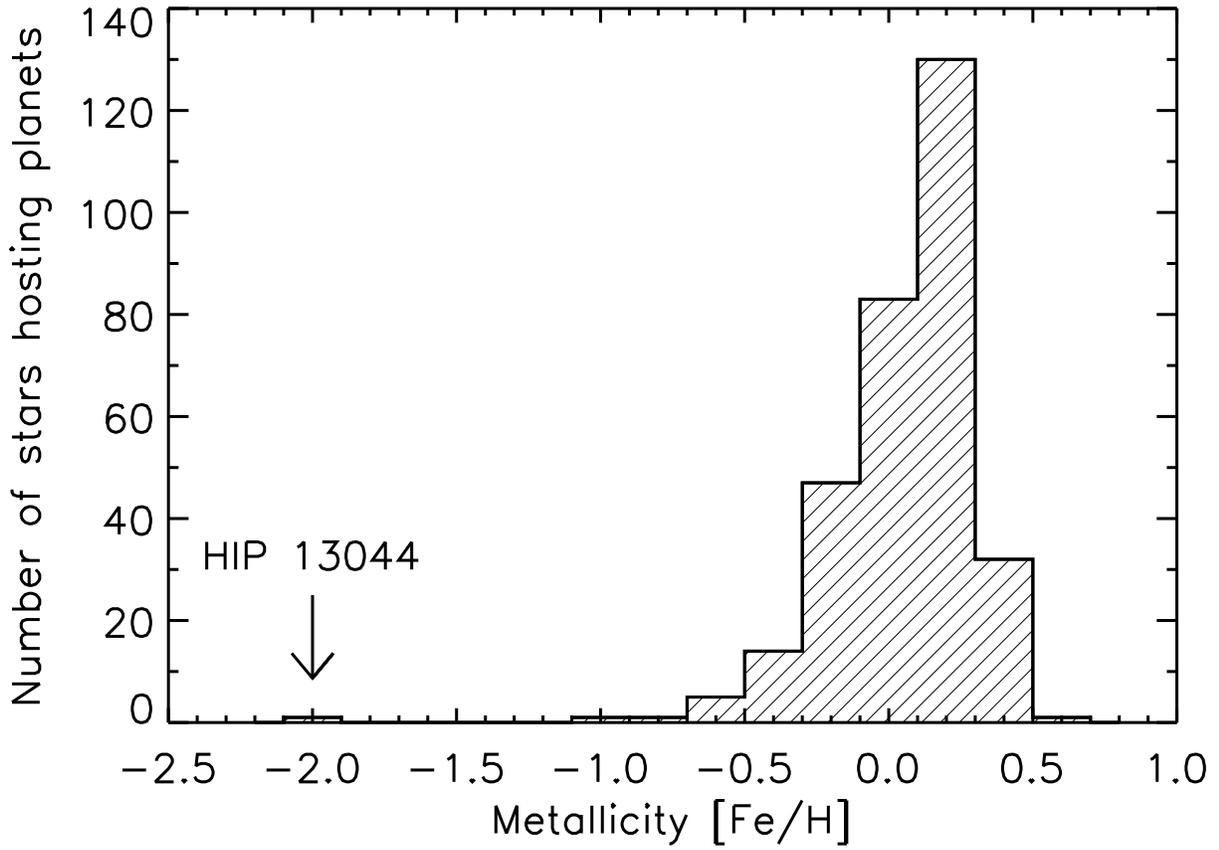}
 \caption{Distribution of the metallicity [Fe/H] of planet-hosting stars.}
 \end{figure*}
 
\clearpage

\section{Supporting online material}
{\small This supplementary online material is part of the article  
with the title {\bf ''A Giant Planet Around a Metal-poor Star of 
Extragalactic Origin"}. 
Here we give additional information about the stellar parameters of \mbox{HIP 13044} 
(Section~\ref{text1}) and more detailed descriptions of the observations 
as well as the stellar radial velocity (RV) computations (Section~\ref{text2}). 
As reported in the manuscript, the RV variation has a period of 16.2 days 
and semi-amplitude of 120 $\mathrm{m\,s}^{-1}$ which most likely 
originate from an unseen planetary companion. 
In the following sections we describe briefly the technique that 
we used to investigate the stellar activity (Section~\ref{text3}).
We analyzed the line profile asymmetry (bisector) and Ca II $\lambda$849.8 
line, which are excellent stellar activity indicators. Additionaly, we also 
present our analysis of the Hipparcos and SuperWASP photometric data.}

\subsection{Stellar parameters\label{text1}}

\par
\mbox{HIP 13044} (CD-36 1052, CPD-36 287) is a star of spectral 
type F2 (SIMBAD). It has a visual magnitude of $m_\mathrm{V}=$9.94 mag 
(Hipparcos) and distance of $\sim$700 pc \citep{car86,car08a}. 
\mbox{HIP 13044} resides in the red part of the Horizontal Branch (RHB), 
which is separated from the blue and extreme Horizontal Branch (BHB and EHB) 
stars by the RR Lyrae instability strip. 

The Hipparcos archive data \citep{per97} revealed variations in the
photometry and astrometry. 
From the Hipparcos intermediate astrometry data, 
we calculated the {\it rms} of the difference between the 
observed great-circle abscissa and 
the solution computed with the reference astrometric parameters 
to be $9\times 10^{-3}$ arcsec. 
While the Hipparcos photometry data shows only a marginal significant 
periodicity at $P=7.1$ hours, photometric observations with SuperWASP \citep{pol06} reveal 
some periodicities with timescales of several hours up to few days 
(Section~\ref{phot}). These signals might correspond to the stellar pulsations. 

As a member of the Helmi stellar stream \citep{hel99}, \mbox{HIP 13044} 
shares the property of other stream members, like the low iron abundances 
([Fe/H]$=-1.8$ for 33 stream members) \citep{kep07,kle09} 
and a chemical similarity to typical inner halo stars \citep{roe10}. 
An extensive abundance analysis of \mbox{HIP 13044} has been also 
presented in \citet{roe10}.

\begin{table}[h]
\begin{center}

{\small Table. S1\\
Stellar parameters of \mbox{HIP 13044}

\vspace{0.2cm}
\begin{tabular}{llll}
\hline
\hline  
  Parameter      &  Value		&  Unit   	& Reference \\
 \hline
 Spectral type      &  F2		&     		& {\sc SIMBAD}\\
 $m_{V}$            &  9.94 		& mag 		& Hipparcos\\
 distance           &  701$\pm$20 	& pc  		& \citet{bee90,car86,car08a}\\
 $T_{\mathrm{eff}}$ &  6025$\pm$63	& K     	& \citet{car08a,roe10}\\
 $R_*$		    &  6.7$\pm$0.3 	& R$_{\odot}$ 	& \citet{car08a}\\

 $\log{g}$	    &  2.69$\pm$0.3	&       	& \citet{car08a}\\
 $m$ 		    &  0.8$\pm$0.1	& M$_{\odot}$ 	& this work\\
 $[Fe/H]$           &  -2.09$\pm$0.26	& [Fe/H]$_{\odot}$ & \citet{chi00,car08b,roe10,bee90}\\
 $v \sin{i}$	    &  8.8$\pm$0.8   	& $\mathrm{km\,s}^{-1}$ & \citet{car08a} \\
 		    &  11.7$\pm$1.0   	& $\mathrm{km\,s}^{-1}$ & this work\\
 \hline 
 \hline
 \end{tabular}}
\end{center}
\end{table}

\begin{table*}[t]
\begin{center}
{\small Table. S2\\
RV measurements of \mbox{HIP 13044}

 \vspace{0.2cm}
 \begin{tabular}{|c|c|c|c|c|c|}
 \hline
 \hline 
{\bf Julian Date}      &  {\bf RV}		&    {\bf $\sigma$} & {\bf Julian Date}      &  {\bf RV}		&    {\bf $\sigma$}\\
 -2450000         &  -300 km/s	&    m/s &  -2450000         &  -300 km/s	&    m/s\\

 \hline 
5103.8853  & 	   4225.32  & 50.50 & 5225.6069  &	 4288.14  & 58.04   \\ 
5106.7933  & 	   4258.74  & 47.50 & 5254.5296  &	 4003.23  & 56.92   \\ 
5164.7565  & 	   4282.91  & 50.86 & 5255.5402  &	 4097.93  & 52.17   \\ 
5168.6543  & 	   4169.13  & 50.22 & 5256.5457  &	 4178.24  & 46.06   \\ 
5170.7212  & 	   4102.30  & 50.42 & 5258.5407  &	 4240.39  & 56.84   \\ 
5172.6968  & 	   4081.84  & 63.61 & 5262.5523  &	 4279.91  & 55.03   \\ 
5174.6528  & 	   4062.56  & 42.69 & 5263.5499  &	 4276.28  & 52.83   \\ 
5175.5576  & 	   4116.08  & 51.69 & 5268.5164  &	 4066.90  & 48.92   \\ 
5197.5703  & 	   4282.29  & 47.90 & 5269.5159  &	 4101.27  & 69.23   \\ 
5199.5236  & 	   4285.80  & 48.63 & 5396.8722  &	 4240.15  & 53.09   \\ 
5199.5666  & 	   4264.87  & 44.12 & 5397.8666  &	 4196.54  & 75.11   \\ 
5200.6345  & 	   4223.49  & 51.08 & 5399.8604  &	 4089.90  & 111.8   \\ 
5202.6003  & 	   4094.38  & 46.72 & 5400.9307  &	 4006.32  & 52.99   \\ 
5203.6050  & 	   4017.79  & 54.59 & 5401.8926  &	 4095.39  & 53.14   \\ 
5204.6531  & 	   4148.22  & 64.52 & 5402.8419  &	 4081.08  & 113.26  \\ 
5223.6119  & 	   4017.52  & 54.59 & 5403.8388  &	 4153.56  & 85.38   \\ 
5224.5837  & 	   4203.70  & 51.66 & 5404.8850  &	 4232.36  & 57.91   \\ 
5225.5329  & 	   4180.37  & 50.41 & 5405.8472  &	 4249.60  & 78.77   \\ 

 \hline 			       
 \hline
 \end{tabular}}
 \end{center}
 \end{table*} 

Table~S1 summarizes the fundamental stellar parameters, such as 
the effective temperature $T_{\mathrm{eff}}$, surface gravity $\log{g}$, 
stellar radius $R_\ast$, stellar mass and metallicity [Fe/H], from the
literature and our spectroscopic measurements.
The stellar mass has been inferred from the knowledge of 
the stellar radius $R_\ast$ and surface gravity $\log{g}$. 
We calculated a stellar mass of $0.8\pm0.1 M_{\odot}$. 
\citet{car08a} have measured $v \sin{i}= 8.8\, \mathrm{km\,s}^{-1}$, 
whereas we obtained $v \sin{i}= 11.7\, \mathrm{km\,s}^{-1}$. 
The discrepancy between the two results is probably 
caused by the different methods used. \cite{car08a} used a 
Fourier transform method, whereas we used a cross-correlation 
technique to measure the $v \sin{i}$ \citep{set04}.
For this work we adopt $v \sin{i}= 10.25\pm2.1 \, \mathrm{km\,s}^{-1}$ 
which is the mean value of both measurements.

\subsection{Radial velocity measurements\label{text2}}

\par
The observations of \mbox{HIP 13044} have been performed with
FEROS, a high-resolution spectrograph ($R$ = 48\,000) attached to the 2.2 meter
MPG/ESO telescope, located at the ESO La Silla observatory. 
The spectrograph is equiped with two fibers and has a wavelength coverage
from 350 to 920\,nm that is divided into 39 echelle orders \citep{kau00}.
The first fiber is used to record the stellar spectrum, whereas the second
fiber can be used for a simultaneous calibration with a ThArNe lamp, in
which the instrumental velocity drift during the night can be corrected.

The raw spectroscopic data have been reduced with an online data
reduction pipeline that produced 39 one-dimensional spectra 
of each fiber. The stellar RVs have been computed by using 
a cross-correlation technique, where the
stellar spectrum is cross-correlated with a numerical template (mask).
The data reduction and RV computation methods have been described in 
\citet{set03}.
From our long-term RV surveys (2003 to 2010) with FEROS,
an accuracy in RV of 11 $\mathrm{m\,s}^{-1}$ has been measured 
for a spectroscopic RV standard star \mbox{HD 10700} (spectral type K8V).

However, this accuracy cannot be achieved for stars with fewer absorption 
lines due to higher effective temperature and for fast rotating stars, 
since the spectral lines are broader.
We obtained a typical accuracy of $\sim50\, \mathrm{m\,s}^{-1}$ for the
individual RV measurements of \mbox{HIP 13044}.
\begin{figure}[t]
\centering
\includegraphics[width=0.9\textwidth]{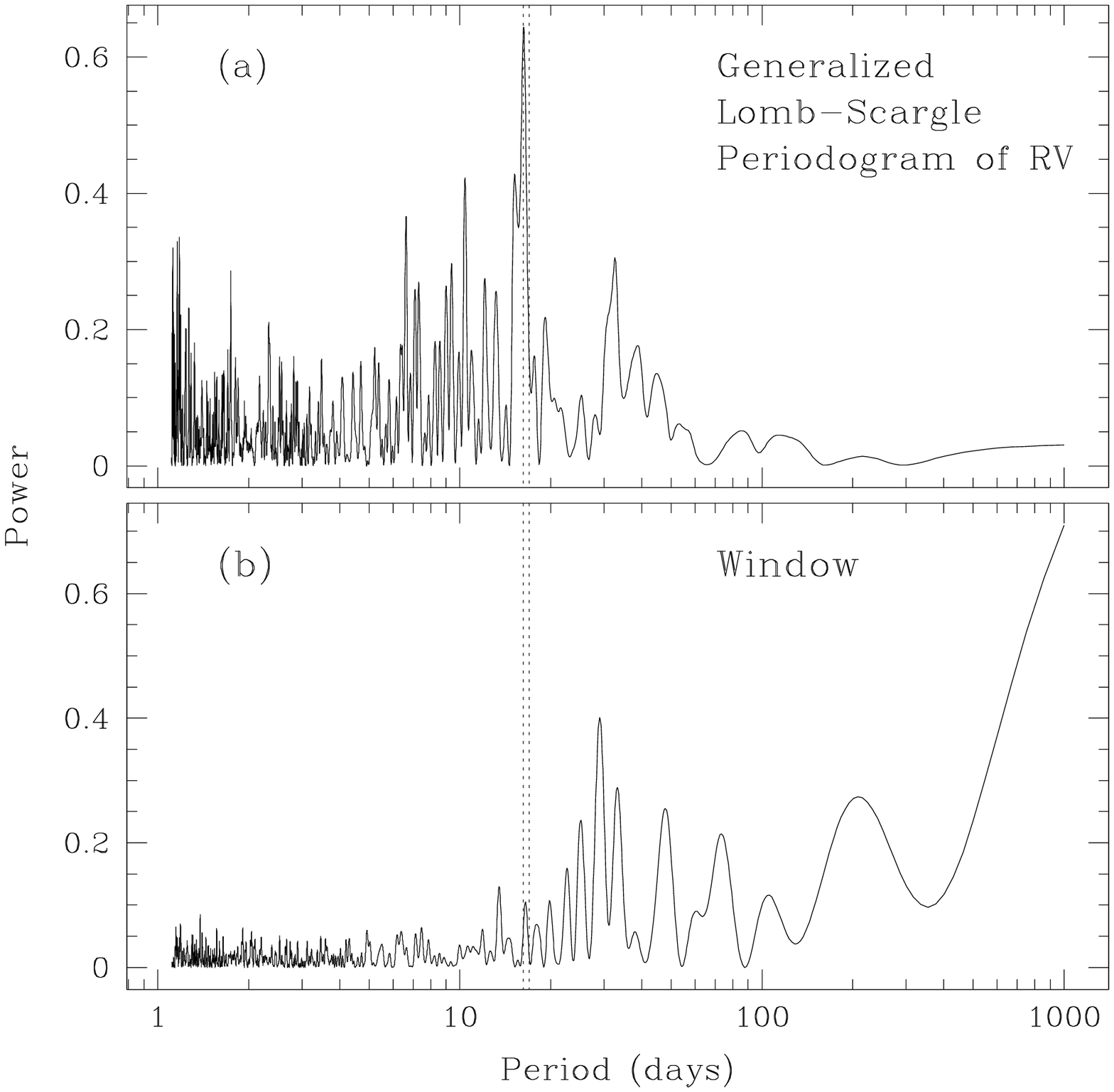} 
\caption{{\small Upper panel: Generalized Lomb-Scargle (GLS) Periodogram of the
RV measurements of \mbox{HIP 13044}. The window function is shown in 
the lower panel. The GLS periodogram shows a significant peak at $\sim$16 days (dotted
lines). The window function does not show any peak around this period.}}
\label{fig:S1}
\end{figure}
In total, 37 high-resolution stellar spectra have been obtained with FEROS. 
However, one spectrum could not be used for the RV analysis due to 
a calibration problem. Thus, we use 36 spectra for our 
spectroscopic study.
For \mbox{HIP 13044} (spectral type F2) we measured the stellar RVs by
computing the cross-correlations of the stellar spectra with a mask,
designed for stars of the spectral type F0. In addition, we also created a 
special cross-correlation mask from the mean spectrum of \mbox{HIP 13044} 
spectra. This mask, called ''HIP13044 temp", 
contains 550 selected spectral lines. To find out if there is inconsistency in
the characteristics, such as period and  amplitude, of the
RV variation, as a result of using two different masks, 
we cross-correlated the spectra of \mbox{HIP 13044} with
this special mask HIP13044-temp. 
Except the zero-point offsets, we found no discrepancies 
in the period and amplitude of the RV variation 
when using either the F0 or the HIP13044-temp masks. 
Thus, in this work we present only the RV measurements with 
the standard F0 mask.

The RV values of \mbox{HIP 13044} have been computed from the mean RVs of 20
usable echelle orders. Table~S2 gives the values of the RV measurements. 
A periodogram analysis by using a Lomb-Scargle (LS) periodogram \citep{sca82}
shows a significant peak at $P\sim$16 days with a False Alarm 
Probability (FAP$_\mathrm{LS}$) of $2\times 10^{-3}$. 
In addition, we also applied the Generalized Lomb-Scargle 
(GLS) periodogram \citep{zec09} on the data. By using GLS we 
found the highest peak at $P=16.23$ days with 
a FAP$_\mathrm{GLS}=5.5\times10^{-6}$ (Fig.~\ref{fig:S1}). 
The RV variation has a semi-amplitude of 120 $\mathrm{m\,s}^{-1}$. 


\subsection{Stellar activity\label{text3}}
Stellar activity is an important issue when interpreting the observed 
RV variations. Stellar rotational modulation is particularly critical 
when probing the planetary hypothesis. 
It can mimic the star's reflex motion from a companion and 
so lead to a wrong interpretation of the RV variation. Another type of 
stellar activity that can produce periodic RV variation is stellar 
pulsation. We investigate both types of stellar activity of \mbox{HIP 13044} 
in the following subsections.

\subsubsection{Rotational modulation}
A first approximation of the stellar rotation period can be obtained from the
knowledge of the projected rotational velocity $v \sin i$ and stellar radius
$R_*$. The maximum rotation period can be calculated from 
$ P_{\mathrm{rot}}/\sin i = 2\pi R_* / v \sin i$ where $i$ 
is the inclination angle of the stellar rotation axis.
With a stellar radius of 6.7 R$_\odot$ and $v \sin{i}$ = 10.25 $\mathrm{km\,s}^{-1}$, 
we obtained a maximum rotation period of $\sim$33 days.
However, since $\sin{i}$ remains unknown, the real rotational period
cannot be determined by this method.
\begin{figure}[t]
\includegraphics[width=0.9\textwidth]{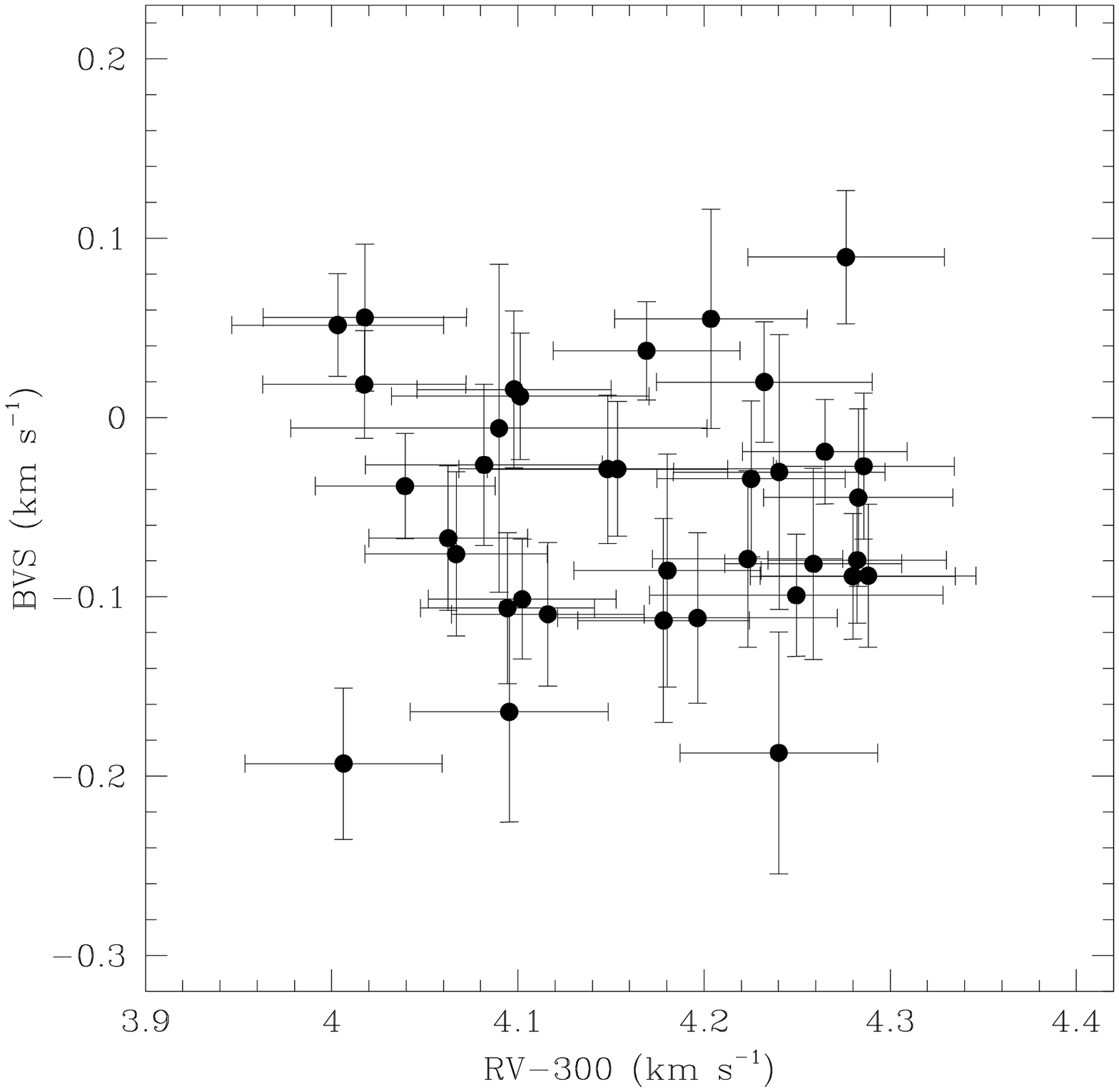}
\caption{{\small The plot of BVS versus RV shows a weak correlation 
between both quantities. With all 36 points we calculated a 
correlation coefficient of -0.13. A carefull rejection of 4 potential 
''outliers" yields a correlation coefficient of -0.4, which is slightly 
higher, but still insignificant.}}
\label{fig:S2}
\end{figure}
\begin{figure}[ht]
\includegraphics[width=0.9\textwidth]{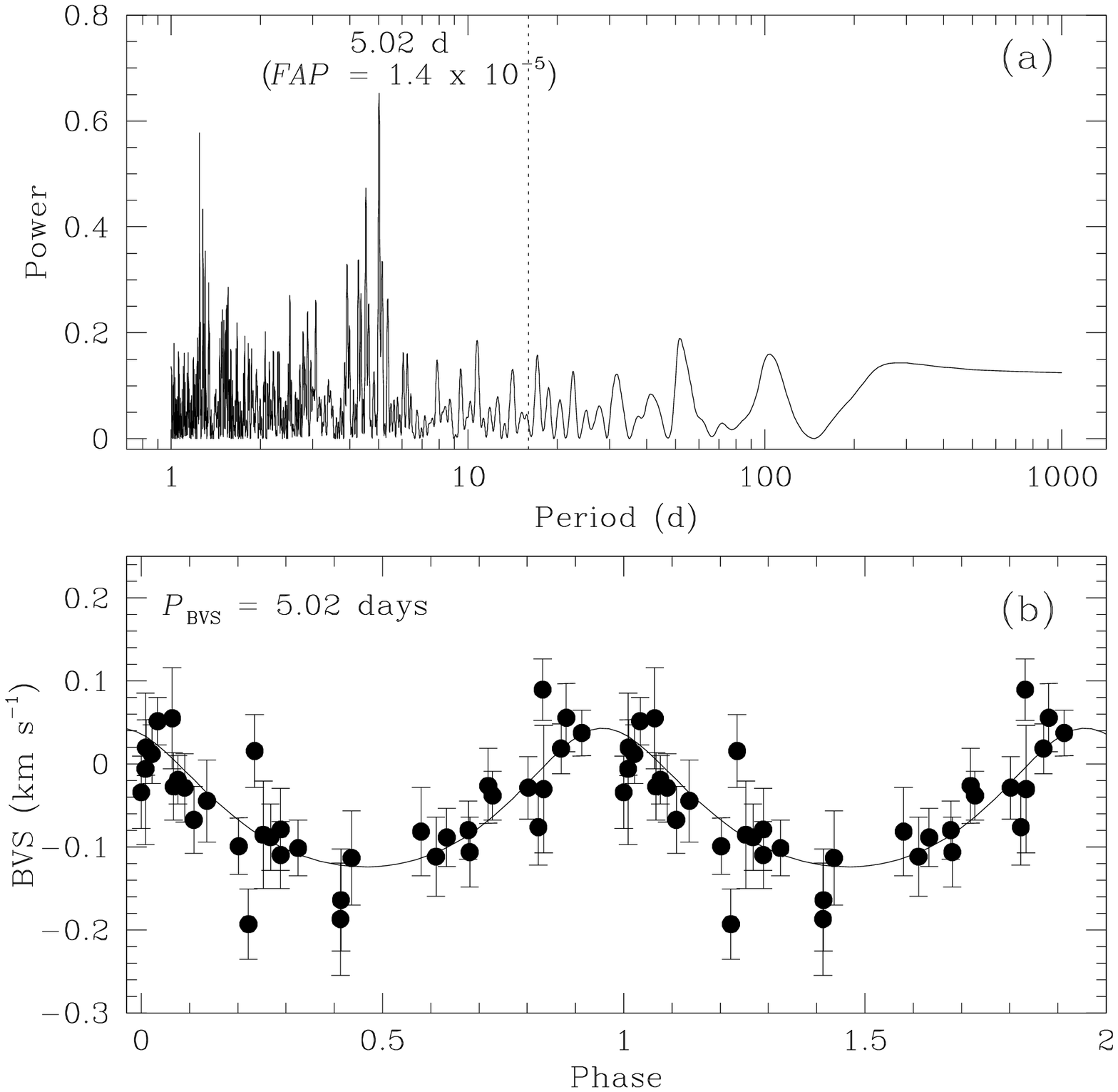}
\caption{{\small Periodogram analysis of the BVS variation showing a periodicity 
of 5.02 days with a FAP of $1.4\times 10^{-5}$ (upper panel). 
The lower panel shows a phase-folded plot of the BVS variation. 
The semi-amplitude of the variation is $\sim70\mathrm{m\,s}^{-1}$.}}
\label{fig:S3}
\end{figure}
Obviously, a better approach should be made to determine the rotation 
period of the star. Photometric observation is a known classical method to 
find stellar rotational modulation. Periodic photometric variation 
due to migrating surface inhomogeneities (starspots) 
can reveal the stellar rotation. 
However, although there are photometric observations by Hipparcos 
and SuperWASP of \mbox{HIP 13044}, the observed variations, even if they 
are periodic, still cannot be directly interpreted as the result of 
the stellar rotational modulation.
For post main-sequence stars like RGB and HB stars, short-period 
variations of hours to few days in the photometry are usually caused 
by the stellar pulsations rather than by rotational modulation.


\subsubsubsection{Bisector\label{sec:bisector}}

We investigated the rotational modulation by examining the variation of 
the line profile asymmetry (bisector). In this work we use the bisector 
velocity span (BVS) which gives the velocity difference between the upper 
and lower part of the absorption line profile \citep[for a 
definition of BVS see e.g.][]{hat96}. We examined the BVS of the stellar spectra and 
searched for evidence for correlation between BVS and RV variation.  
As shown in Fig.~\ref{fig:S2} we found only a weak correlation between BVS and RV 
(correlation coefficient $c= -0.13$). 
Since the correlation coefficient is sensitive to the potential outliers 
in the data, we have carefully investigated this factor by removing 
few ``outliers'' in Fig.~\ref{fig:S2}. 
The highest correlation coefficient we obtained after removing 4 outliers 
is $c= -0.4$. 
By bootstrapping the data 100,000 times, we derived a standard deviation of 0.16 for $c$. Together with the 
original value ($c= -0.13$), this confirms the weakness of the correlation 
between BVS and RV.

Interestingly, we found that the BVS variation 
shows a clear periodicity of 5.02 days. 
The FAP$_\mathrm{GLS}$ of this period is $1.4\times10^{-5}$. 
Fig.~\ref{fig:S3} shows the periodogram of BVS and the phase-folded 
BVS variation. Following the idea that migrating starspots on the stellar surface 
cause the variations in the line profile asymmetry or BVS, which is an effect of  
the rotational modulation, the period of $P_\mathrm{BVS}=$5.02 days is a possible 
candidate for the stellar rotation period.
In particular, we found no period around 16 days in the
BVS variation. Hence, for \mbox{HIP 13044} the period of the RV variation 
can be distinguished clearly from the anticipated stellar rotation period. 

\subsubsubsection{Ca II $\lambda8498$ equivalent width\label{ew}}

The emission cores in Ca II K ($\lambda$393.4) and H ($\lambda$396.7) 
as well as the Ca II IR triplet lines ($\lambda 849.8,854.2,866.2$) are 
well known stellar activity indicators.
However, for \mbox{HIP 13044} we found no periodic variation 
of the Ca II K lines. We did not use the Ca II H line for our analysis 
since it is blended by the H $\epsilon$ line. 

\begin{figure}[t]
\centering
\includegraphics[width=0.9\textwidth]{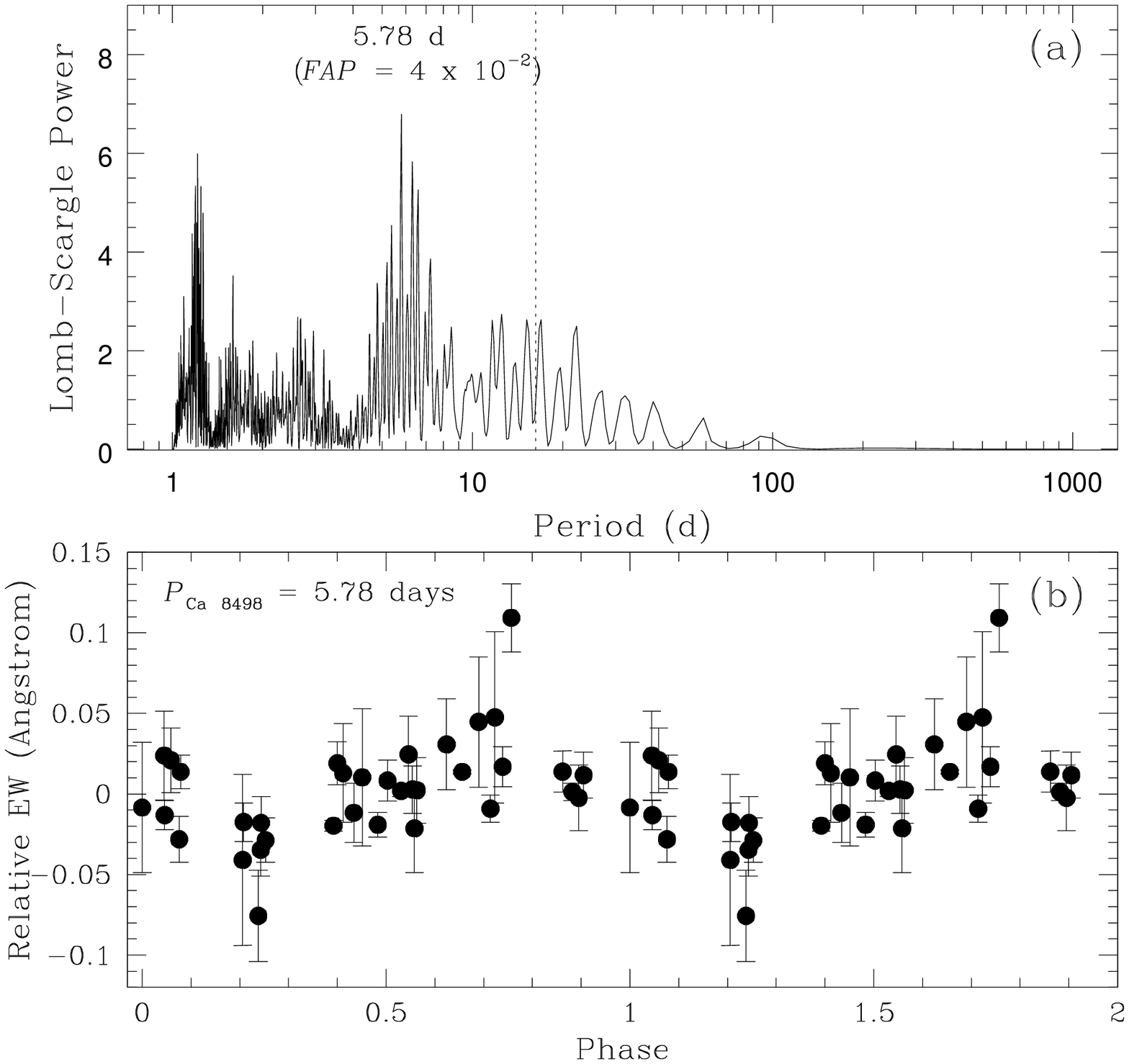}
\caption{{\small Analysis of the CaII $\lambda$849.8 EW variation
by using a Lomb-Scargle periodogram.  The LS-periodogram 
shows a peak at 5.78 days with a marginal FAP of 4\%. 
The lower panel is the plot of the EW variation of CaII $\lambda$849.8 
that is phase-folded with a period of 5.78 days. }}
\label{fig:S4}
\end{figure}

\begin{figure}[t]
\centering
\includegraphics[width=0.9\textwidth]{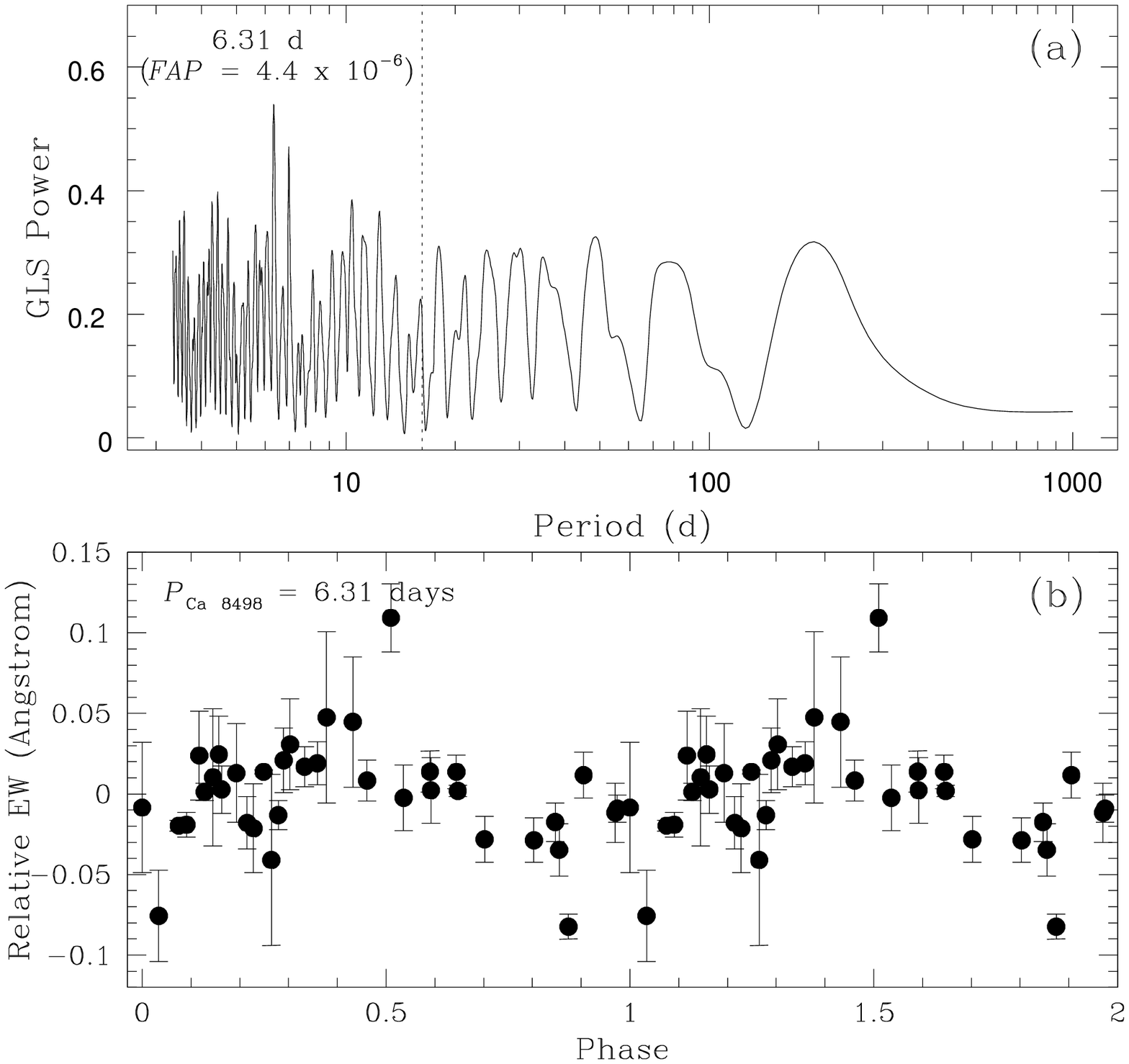}
\caption{{\small Analysis of the CaII $\lambda$849.8 EW variation
by using the Generalized Lomb-Scargle periodogram. The period 
found in the GLS-periodogram is only slightly different from 
that obtained from the LS-periodogram. The GLS-periodogram shows a peak at 6.31 days 
with a FAP of $4.4\times10^{-6}$. The phase-folded plot 
with $P=6.31$ days is shown in the lower panel.}}
\label{fig:S5}
\end{figure}

Furthermore, we measured the equivalent width (EW) of Ca II IR triplets. 
While we did not find any periodicities in Ca II $\lambda 854.2$ and $\lambda
866.2$, we found periodic EW variations in Ca II $\lambda$849.8. 
The LS periodogram shows a peak at 5.78 days with a marginal 
FAP$_\mathrm{LS}$ of $\sim$4\% (Fig.~\ref{fig:S4}), 
whereas the GLS periodogram shows a period of 6.31 days 
(FAP$_\mathrm{GLS}=4.4\times 10^{-6}$), as shown in Fig.~\ref{fig:S5}.
The small discprepancy between the two periods could result 
from the different fitting and weighting methods used in both periodograms.
We conclude that the activity indicator Ca II 849.8 has a period of 
$6.05\pm0.3$ days.

The results from BVS and Ca II $\lambda$8498 EW variation lead to the conclusion 
that the stellar rotation period lies between 5 and 6 days. 
Finally, we adopted $P=5.53\pm 0.73$ days, i.e., the mean value of the periods 
of both spectroscopic activity indicators as the rotation period of \mbox{HIP 13044}.

\subsubsection{Stellar pulsations}

\subsubsubsection {Pulsations of Horizontal Branch stars}
Radial and non-radial stellar pulsations have been observed in post 
main-sequence stars, such as RGB, AGB stars and white dwarfs. 
Indeed, many HB stars 
are also known to be pulsators. However, until now there are not many 
studies or reported detections of RHB stellar pulsations.
Theoretical models of pulsations of red variable HB stars by \citet{xio98} 
predicted that the oscillation periods of high-order overtones are of the 
order of a few tenths of a day (few hours).

Furthermore, the oscillation frequency $\nu_\mathrm{max}$ 
can be predicted from an empirical relation by \cite{kje95}. 
For \mbox{HIP 13044} we calculated $\nu_\mathrm{max}\approx 53\mu$Hz 
which corresponds to a pulsation period of $\sim$5.2 hours.
This hypothesis has some support from the studies of HB stars 
in metal-poor globular clusters, e.g., NGC 6397 \citep{ste09}.

\subsubsubsection {RR Lyrae variables}
Since the location of \mbox{HIP 13044} in the Horizontal Branch is close to the RR 
Lyrae instability regime, it is also important to investigate the pulsations 
of those stars and compare them to the detected RV periodicity in \mbox{HIP 13044}. 
\citet{har05} observed photometric variability in the metal-poor cluster M3. 
They found 180 variable stars with periods within 1 day. 
Similar observations, but for M15 by \citet{cor08} showed that the RR 
Lyrae variables in M15 have pulsation periods between 0.06 and 1.44 days.
These results show that a period of $\sim$16 days is not likely for oscillation 
characteristics of RR Lyrae stars.

\subsubsubsection {W Virginis variables}
W Virginis variable stars belong to Population II Cepheids that have 
pulsation periods of 10 to 30 days. Thus, there might be a concern about 
the $\sim$16 day periodicity in \mbox{HIP 13044} being due to pulsations 
rather than a planetary companion.  
However, this requires that \mbox{HIP 13044} itself must belong to W Virginis
variables. 

Since for W Virginis stars, similar to the classical Cepheids there exist 
Period-Luminosity relations \citep[e.g.][]{cox79}, one can calculate the 
expected stellar luminosity from the oscillation period.
\citet{wal84} listed some basic data of the Population II Cepheids 
in globular clusters. For Population II Cepheid stars with periodicities 
of $\sim$16 days in the globular clusters M3, M10, M2 and $\omega$ Cen, 
their absolute magnitudes $M_V$ are between -2.2 and -2.7 mag. 
This is about 3 magnitudes higher than the absolute magnitude of \mbox{HIP 13044} 
($M_V\sim0.6$). Less luminous Population II Cepheid stars 
with $M_V$ between 0.1 and 0.6 are expected to exhibit pulsations 
with periods of 1--2 days.    

Additionaly, according to \citet{pri03} one can calculate the absolute 
magnitude of the star, if it indeed belongs to W Virginis variables. 
For $P=16.2$ days, the expected absolute magnitude $M_V$ is -1.925 mag. 
Because of the lack of supporting W Virginis characteristics, we concluded 
that \mbox{HIP 13044} does not belong to the W Virginis variables. 
The 16 day periodicity is, therefore, not related to the stellar oscillations.

\subsubsubsection {Photometric data\label{phot}}

We analyzed the Hipparcos photometry data 
to investigate the effect of the stellar pulsations. 
168 photometric measurements of \mbox{HIP 13044} are available in 
the public archive. The LS periodogram of the Hipparcos data 
shows no significant signal. The highest peak corresponds to a period of 7.1 hours 
with a marginal FAP=1.8\%. If this period is true, it would be in agreement with the 
predicted pulsation frequency \citep{kje95}. 
The Hipparcos data is, however, not sufficient to derive more information 
about the stellar pulsations.

\begin{figure*}[t]
\center
\includegraphics[width=11cm,height=11cm]{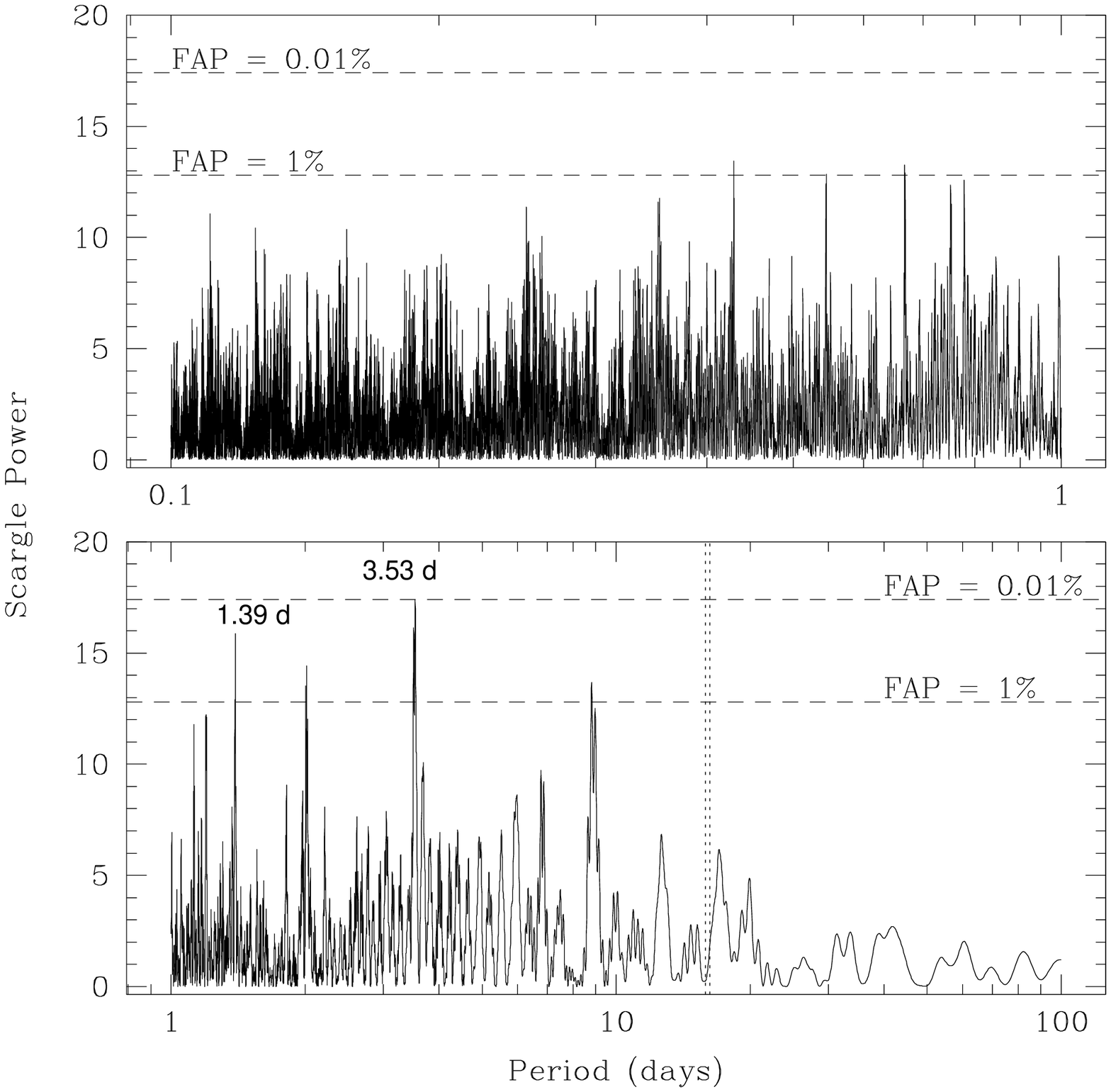}
\caption{{\small Lomb-Scargle periodogram of the SuperWASP photometry data. 
The periodogram shows only marginal significant (FAP$\sim$1\%) peaks that 
correspond to intra-day periodicities (upper panel). It is not clear, 
if these signals are related to the stellar oscillations predicted by the theory. 
Nevertheless, in the lower panel the periodogram shows two significant peaks 
at 3.53 days (FAP=$2\times10^{-4}$) and 1.39 days (FAP=$5\times10^{-4}$). 
These signals are probably harmonic to each other. 
In this extensive data set no photometric variation with 
a period of $\sim$16 days has been detected (dotted lines).}}
\label{fig:S6}
\end{figure*}

In addition, we also analyzed the photometric observations taken by 
the SuperWASP project \cite{pol06}. 
The observations were carried out in 2 blocks, each of 3--4 months. 
For \mbox{HIP 13044} there are 3620 photometric measurements available in the archive.
After removing 10 outliers, probably caused by systematic errors, we used 3610 
data points for the periodogram analysis. 
With such a large number of measurements, a possible 16-d periodicity in 
the photometric variations should be easily detectable.

In Fig.~\ref{fig:S6} we present the LS-periodogram of the 
SuperWASP data. The LS-periodogram shows few marginal significant 
(FAP$\sim$1\%) intra-day periodicities. However, it is not clear, 
if these signals can be attributed to the stellar oscillations as predicted 
by the theory. Interestingly, we observe significant peaks at 3.5 days 
(FAP=$2\times 10^{-4}$) and 1.4 days (FAP=$5\times 10^{-4}$). 
The two signals are most likely harmonic to each other. 
Since the FAPs of both signals are of the same order ($10^{-4}$), 
the determination of the exact value of the period is still difficult.
It is also possible that these periods correspond to oscillation overtones of very high order. 
Most important, however, is the fact that no signal at 16.2 days is detected in the LS-periodogram 
of the SuperWASP photometry data. 
Due to the absence of a periodic photometric variation of 16.2 days, 
stellar pulsations are not likely to be the source of 
the observed periodic RV variation. 

\subsection{Concluding remarks}
Based on the spectroscopic and photometric analysis described above, 
stellar rotational modulation and pulsations can be ruled out as 
the source of the periodic RV variation.
The spectroscopic stellar activity indicators show a period of 5--6 days 
(bisectors and Ca II $\lambda849.8$) whereas the photometric data shows 
periodicities of intra-day to few days, 
that can be clearly distinguished from the period of the RV variation. 
We conclude that the observed 16.2 days RV variation cannot be due  
to rotational modulation or stellar pulsations, but is rather caused by 
the presence of an unseen planetary companion. 

\section*{Acknowledgements}
We thank our MPIA colleagues: W. Wang, C. Brasseur, R. Lachaume, 
M. Zechmeister and D. Fedele for the spectroscopic observations
with FEROS. We also thank Dr. M. Perryman, Dr. E. Bear, Dr.
N. Soker and Dr. P. Maxted for the fruitful discussion, comments and suggestions 
that helped to improve this paper.

\clearpage

\end{document}